\documentclass[aps,superscriptaddress]{revtex4}

\usepackage{xcolor}
\definecolor{tab20green}{rgb}{0.1725490196,0.6274509804,0.1725490196}
\definecolor{tab20red}{rgb}{0.8392156863,0.1529411765,0.1568627451}
\definecolor{tab20blue}{rgb}{0.1215686275,0.4666666667,0.7058823529}
\usepackage{amsfonts,amssymb,amsmath} 
\usepackage{graphicx}
\usepackage{dsfont}
\usepackage{soul}
\usepackage{ulem}
\usepackage[
  colorlinks,
  linkcolor=blue,
  urlcolor=blue,
  citecolor=blue,
  plainpages=false,
  pdfpagelabels,
]{hyperref}

\graphicspath{{./figs/}}

\DeclareMathAlphabet\boldsymbolcal{OMS}{cmsy}{b}{n}

\newcommand{\baleqn}{\begin{equation}\begin{aligned}[b]}
\newcommand{\ealeqn}{\end{aligned}\end{equation}}
\newcommand{\baleqns}{\begin{equation*}\begin{aligned}}
\newcommand{\ealeqns}{\end{aligned}\end{equation*}}

\newcommand\bal{\begin{aligned}}
\newcommand\eal{\end{aligned}}
\newcommand{\bq}{\begin{eqnarray}}
\newcommand{\eq}{\end{eqnarray}}

\newcommand{\beq}{\begin{equation}}
\newcommand{\eeq}{\end{equation}}
\newcommand{\beqa}{\begin{eqnarray}}
\newcommand{\eeqa}{\end{eqnarray}}

\begin{document}
\title{Effective field theories for interacting boundaries of 3D topological crystalline insulators through bosonisation}

\author{Patricio Salgado-Rebolledo}
\affiliation{School of Physics and Astronomy, University of Leeds, Leeds, LS2 9JT, United Kingdom}
\author{Giandomenico Palumbo}
\affiliation{Center for Nonlinear Phenomena and Complex Systems,
Universit\'e Libre de Bruxelles, CP 231, Campus Plaine, B-1050 Brussels, Belgium}
\author{Jiannis K. Pachos}
\affiliation{School of Physics and Astronomy, University of Leeds, Leeds, LS2 9JT, United Kingdom}

\date{\today \;\;e-mail: psalgadoreb@gmail.com}

\begin{abstract}
Here, we analyse two Dirac fermion species in two spatial dimensions in the presence of general quartic contact interactions. By employing functional bosonisation techniques, we demonstrate that depending on the couplings of the fermion interactions the system can be effectively described by a rich variety of topologically massive gauge theories. Among these effective theories, we obtain an extended Chern-Simons theory with higher order derivatives as well as two coupled Chern-Simons theories. Our formalism allows for a general description of interacting fermions emerging, for example, at the gapped boundary of three-dimensional topological crystalline insulators. 
\end{abstract}

\maketitle

\section{Introduction}

Time-reversal-invariant topological insulators are among the most well studied topological phases of matter. In three dimensions, they are characterised by suitable topological numbers in the bulk that guarantee the existence of topologically protected massless Dirac fermions on the boundary \cite{moore2010birth,Qi:2011zya}.
Although the topological invariant is a $\mathbb{Z}_2$ number, on the slab geometry, it has been shown that robust surface states are given by an odd number of Dirac fermions per boundary \cite{Fu2}. The situation changes in the case of three-dimensional topological crystalline insulators (TCIs), namely topological insulators characterised by further crystalline symmetries, such as mirror and rotation symmetries \cite{Fu,Hsieh,Ryu,slager2013space,Furusaki,hsieh2012topological,tanaka2012experimental,Shiozaki,PhysRevX.7.041069}.
In particular, for three-dimensional TCIs protected by a single mirror symmetry,
one can define the so called mirror
Chern number $n_M$ on a given two-dimensional plane,
which is invariant under the mirror symmetry. These phases host $n = |n_M|$ Dirac cones on each boundary \cite{Hsieh}. 
Recently, mirror-invariant boundary interactions in these systems have been intensively studied by employing several approaches, such as non-linear sigma models \cite{Morimoto,Song}, the coupled-wire method for $n_M=2$ \cite{Hong}, Higgs phases for $n_M=4$ \cite{Qi} and symmetry arguments for $n_M=8$ \cite{Isobe}.

Bosonisation represents another important quantum-field-theory approach to study interacting Dirac fermions. It was originally formulated in 1+1 dimensions to map the the massive Thirring model to the Sine-Gordon theory \cite{Coleman,Maldestam} and then extended in higher-dimensional relativistic systems under the name of 
functional bosonisation \cite{Fradkin:1994tt}. Although this method has numerous implications that are relevant to condensed matter physics, it has been mainly employed in interacting systems involving a single emergent gauge field.

The goal of this work is to analyse gapped and mirror-broken boundary states in presence of quartic contact interactions between several pieces of fermions. 
We assume that the interactions are exclusively acting on the boundary, while the bulk of the system is descried by free topological insulating phase.
 This is similar to the case of 2D time-reversal-invariant topological insulators, where the helical Luttinger liquids appear on the interacting boundary of the system while the 2D bulk states are still related to the free-fermion models \cite{Wu}.
We introduce then an external magnetic field orthogonal to the surface to induce a Dirac mass that breaks both time-reversal and mirror symmetries and consider generic intra- and inter-species interactions. For simplicity, we fix $n_M=2$, as in  \cite{Hong}, and employ functional bosonisation. This approach will allow us to map the self-interacting fermion model to free bosonic models. We are interested in obtaining the low energy topological properties of these effective bosonic models for various configuration of inter and intra-species interactions of the original fermionic model.

Our analysis shows that all the resulting effective models contain topological Chern-Simons terms that usually emerge in a variety of $T$-broken systems such as the quantum Hall states \cite{Zhang:1988wy,Read:1988mp}, surface states of three-dimensional topological insulators \cite{Qi:2011zya} and graphene coupled to external magnetic fields \cite{Fialkovsky:2009wm}.
However, differently from these previous works, 
 we show the existence of a new exotic phase, characterised by a higher-derivative Chern-Simons term \cite{Deser:1999pa}, when one of the intra-species interaction is switched off. This phase supports a massive U(1) boson and a ghost mode, which is completed decoupled from the bosonic mode, and thus it is a ``good ghost" \cite{Hawking:2001yt,Kaparulin:2014vpa}. Moreover, we show the existence of another phase in which the bosonic theory comprises two massive $U(1)\times U(1)$ bosons, with a mutual Chern-Simons term, which generalises the well-know Chern-Simons-Maxwell theory to multi-field gauge fields. The topological sector of this phase resembles the effective action studied in Ref.~\cite{Moore} in the context of thin-film topological insulators.
Importantly, our approach is quite general and can be directly extended to $n_M>2$. This will allow to identify novel topological crystalline phases in presence of very general contact interactions.


\section{Two-fermion interacting system}

The starting point of our construction is a $(2+1)$-dimensional system of two interacting fermion species $\psi$ and $\chi$ living on the boundary of 3D topological crystalline insulator with bulk mirror Chern number $n_M=2$. The corresponding effective action is given by
\beq
\label{Action}
\bal
S[\chi,\psi]=&\int d^3 x \Bigg[ 
 \bar\chi \left(i \gamma^\mu \partial_\mu+m\right)\chi
+\bar\psi\left(i\gamma^\mu \partial_\mu+m\right)\psi
+\frac{V_\chi}{2} \bar\chi\gamma^\mu\chi\bar\chi\gamma_\mu\chi
+V_{\chi\psi}\, \bar\chi\gamma^\mu\chi \bar\psi\gamma_\mu\psi
+\frac{V_\psi}{2}  \bar\psi\gamma^\mu\psi\bar\psi\gamma_\mu\psi
\Bigg],
\eal
\eeq
where $m= B_z \sigma_3$ is the time-reversal broken mass induced by an external magnetic field $B_z$ orthogonal to the surface of the 3D TCI defined on the $xy$-plane.
Here, we use the convention for the Minkowski metric $\eta_{\mu\nu}={\rm diag}(-,+,+)$. The gamma matrices are defined in terms of the Pauli matrices as $\gamma^0=\sigma_3$, $\gamma^1=i\sigma_1$,$\gamma^2=i\sigma_2$ and $\bar\psi=\psi^\dagger \gamma^0$, the Dirac conjugate is $\bar\psi=\psi^\dagger \gamma^0$, and the Clifford algebra has the form $\left\{\gamma^\mu, \gamma ^\nu\right\}=-2  \eta^{\mu\nu}\mathbb{I}_{2\times2}$.
For convenience in the presentation we choose the intra-species coupling constants to be given by $V_\chi=e_\chi^2$ and $V_\psi= e_\psi^2+\xi\alpha^2$, where $e_\chi$, $e_\psi$ and $\alpha$ are real constants, and $\xi=\pm1$, while $V_{\chi\psi} = e_\chi e_\psi$ is the inter-species coupling constants. 

To analytically determine the behaviour of this interaction system we employ functional bosonisation.
This is a powerful approach that will allows us to identify the equivalent bosonic theory describing our model in the low-energy regime.
By defining $k^\mu=\bar\chi\gamma^\mu \chi$, $j^\mu=\bar\psi\gamma^\mu \psi$, the corresponding generating functional has the form
\beq\label{Zeta}
\bal
Z&=\int \mathcal D\bar\chi \mathcal D\chi \mathcal D\bar\psi \mathcal D\psi
\,\exp\bigg\{i\int d^3 x\bigg[ \bar\chi \left(i \gamma^\mu \partial_\mu+m\right)\chi
+\bar\psi\left(i \gamma^\mu \partial_\mu+m\right)\psi \\
& \hskip5truecm+\frac{ 1}{2} \left(e_\chi k^\mu+e_\psi j^\mu \right) \left(e_\chi k_\mu+e_\psi j_\mu \right) 
+\frac{\xi \alpha^2}{2}j^\mu j_\mu 
\bigg]\Bigg\}.
\eal
\eeq
In order to integrate out the fermion field $\chi$, we follow \cite{Fradkin:1994tt,Palumbo:2013rb} (see also \cite{Banerjee:1995ry,Banerjee:1995xm,Santos:2019dlr}) and express the third term in the action as
\beq
\label{gaussint}
\exp \Bigg\{\frac{ i}{2}\int d^3x  \left(e_\chi k^\mu+e_\psi j^\mu \right)\left(e_\chi k_\mu+e_\psi j_\mu \right)\Bigg\} =\int \mathcal D a\, \exp\Bigg\{i\int d^3 x \bigg[- \frac{1}{2}a^\mu a_\mu +   a^\mu  \left(e_\chi k^\mu+e_\psi j^\mu \right)\bigg]\Bigg\},
\eeq
where $a_\mu$ is an Hubbard-Stratonovich vector field.
By replacing this back into the generating functional $Z$, we obtain
\beq
\bal
Z&=\int \mathcal D\bar\chi \mathcal D\chi \mathcal D\bar\psi \mathcal D\psi
\mathcal D a\,\exp\bigg\{i\int d^3 x\bigg[ \bar\chi \left( \gamma^\mu (i\partial_\mu+ e_\chi a_\mu)+m\right)\chi
 \\
&\hskip5.5truecm+\bar\psi\left(\gamma^\mu (i \partial_\mu+ e_\psi a_\mu)+m\right)\psi+ \frac{ \xi \alpha^2}{2}j^\mu j_\mu-\frac{1}{2}a^\mu a_\mu  \bigg]\Bigg\}.
\eal
\eeq
We now integrate out $\chi$ to obtain an effective bosonic action $\Gamma[a]$. In the large mass limit, it can be approximated as \cite{Redlich:1983kn,Redlich:1983dv,Niemi:1983rq}
\beq\label{logdet}
\Gamma[a]=-i\log\left[\det\left( \gamma^\mu (i\partial_\mu+ e_\chi a_\mu)+m\right)\right]\approx \frac{  s_m  e_\chi^2}{8\pi}\int d^3 x \epsilon^{\mu\nu\rho}a_\mu \partial _\nu a_\rho,
\eeq
where $s_m=\frac{ m }{| m |}={\rm sign}(m)$ and  $\epsilon^{\mu\nu\rho}$ is the (2+1)-dimensional Levi-Civita symbol with $\epsilon^{012}=1$.
Therefore we can write
\beq\label{ZSeff}
Z=\int \mathcal D\bar\psi \mathcal D\psi
\mathcal D a  \,\exp \left\{i S_{\rm eff}[\psi,a]\right\},
\eeq
where
\beq
\label{Seff}
S_{\rm eff}[\psi,a]=\int d^3 x \Bigg[-\frac{1}{2}a^\mu a_\mu
+ \frac{ s_m   e_\chi^2}{8\pi} \epsilon^{\mu\nu\rho}a_\mu \partial _\nu a_\rho  
+\bar\psi\left(\gamma^\mu (i\partial_\mu+ e_\psi a_\mu)+m\right)\psi
+\frac{ \xi \alpha^2}{2}j^\mu j_\mu \Bigg].
\eeq
The action \eqref{Seff} holds for general values of the parameters $e_\chi$, $e_\psi$, $\alpha$ and $\xi$. The first two terms are purely  given in terms of the vector field $a_\mu$. They correspond to the self dual action resulting from a single fermionic species, $\chi$, introduced in \cite{Deser:1984kw,Townsend:1983xs}. The total action $S_{\rm eff}$ is an extension to that self-dual model having the field $a_\mu$ coupled to a self-interacting fermionic field $\psi$. In the following we consider specific configurations of these couplings and extract the behaviour of the model in each case. 

\section{Pauli term and Higher-derivative Chern-Simons action}

In this section we show that for a particular value of the couplings the fermionic system \eqref{Action} can be described in the low energy limit by a single fermion field non-minimally coupled to an effective U(1) gauge field. In particular, this coupling configuration gives rise to a higher-derivative Chern-Simons theory \cite{Deser:1999pa}. We start our analysis of \eqref{Seff} by considering the interpolating action
\beq\label{Linterp}
\mathcal S_I[\psi,a,A] =\int d^3x \Bigg[-\frac{1}{2}a^\mu a_\mu +\epsilon^{\mu\nu\rho}a_\mu\partial_\nu A_\rho -\frac{2\pi  s_m }{e_\chi^2}\epsilon^{\mu\nu\rho}A_\mu\partial_\nu A_\rho +e_\psi a^\mu j_\mu +\bar\psi\left(i\gamma^\mu \partial_\mu+m\right)\psi+\frac{\xi \alpha^2}{2} j^\mu j_\mu\Bigg],
\eeq
which is given in terms of the Dirac fermion field $\psi$, the vector field $a_\mu$ and a new gauge field $A_\mu$.
The path integral of $\mathcal S_I[\psi,a,A]$ is equivalent to the functional integral associated to the effective action \eqref{Seff}. By integrating out the field $A_\mu$ in \eqref{Linterp}, we obtain
\beq
\bal
Z_I=\int \mathcal D\bar\psi \mathcal D\psi \mathcal D a \mathcal D A \exp \left\{iS_I[\psi,a,A]\right\} = \int \mathcal D\bar\psi \mathcal D\psi
\mathcal D a  \,\exp \left\{i S_{\rm eff}[\psi,a]
\right\},
\eal
\eeq
where $S_{\rm eff}$ is given by \eqref{Seff}.
On the other hand, by integrating out the vector field $a_\mu$ in the interpolating action \eqref{Linterp} we find
\beq
Z_I=   \int \mathcal D\bar\psi \mathcal D\psi
\mathcal D A  \,\exp \left\{i S^{\rm dual}_{\rm eff}[\psi,A]
\right\},
\eeq
where the action $S^{\rm dual}_{\rm eff}[\psi,A]$ is given by
\beq
\label{Seff22}
\bal
S^{\rm dual}_{\rm eff}[\psi,A]&=\int d^3 x \Bigg[ -\frac{1}{4}F_{\mu\nu}F^{\mu\nu} -\frac{2\pi s_m}{e_\chi^2}\epsilon^{\mu\nu\rho}A_\mu \partial_\nu A_\rho\\
&\hskip 1.5truecm 
+\bar\psi\left(i\gamma^\mu \partial_\mu+m\right)\psi 
+e_\psi \bar\psi \sigma_{ \mu\nu}F^{\mu\nu}\psi
+\frac{1}{2}\left(e_\psi^2+ \xi\alpha^2\right)j^\mu j_\mu \Bigg].
\eal
\eeq
In this dual effective action $S^{\rm dual}_{\rm eff}[\psi,A]$ the field $F_{\mu\nu}=\partial_\mu A_\nu -\partial_\nu A_\mu$ is the field strength associated to the gauge potential $A_\mu$ and we have also used the standard definition $\sigma_{\mu\nu}=\frac{i}{4}[\gamma_\mu,\gamma_\nu]=\frac{1}{2}\epsilon_{\mu\nu\rho}\gamma^\rho$. 

The action $S^{\rm dual}_{\rm eff}[\psi,A]$ is dual to $S_{\rm eff}[\psi,a]$, so it faithfully describes the original system \eqref{Action}. The advantage of $S^{\rm dual}_{\rm eff}[\psi,A]$ is that it is given in terms of the gauge field $A_\mu$ rather than the vector field $a_\mu$ and thus it is easier to identify its topological character. From \eqref{Seff22} we observe that the effective action \eqref{Seff} can be dualised to a Chern-Simons-Maxwell model coupled to the fermion field $\psi$ by means of the Pauli term. Note that, after using the interpolating action, the coupling of the self interaction for $\psi$ has been shifted back to the original value it had in \eqref{Action}. By directly comparing \eqref{Action} and \eqref{Seff22}, we see that we can interpret the Pauli coupling as the low energy description of the mixed interaction term $V_{\chi\psi}\bar\psi\gamma^\mu\psi\bar\chi\gamma_\mu\chi$. The duality between \eqref{Seff22} and \eqref{Seff} has been previously established on-shell by eliminating the field $a_\mu$ or $A_\mu $ from the interpolating action \eqref{Linterp} by means of their corresponding field equations \cite{Gomes:1997mf,Anacleto:2001rp}.

\subsection{Higher derivative Chern-Simons theory: the $e_\psi^2+\xi\alpha^2=0$ case}

We now consider the action \eqref{Seff} for the case where $\alpha^2 = e_\psi^2\neq 0$ and $\xi=-1$, which corresponds to $V_\psi=0$ in \eqref{Action}. In that case, the action \eqref{Seff} does not describe free fermions so that they cannot be integrated out. Moreover for $\xi=-1$ we cannot employ a similar relation to \eqref{gaussint} in order to linearise the interactions. In this case the self interaction in the dual action \eqref{Seff22} vanishes and the effective theory takes the form of a fermion non-minimally coupled to the field strength $F_{\mu\nu}$ by means of the Pauli term. Defining the Hodge dual of the curvature $F_\mu=\frac{1}{2}\epsilon_{\mu\nu\rho}F^{\nu\rho}$ the action takes the form
\beq
\label{Seff21}
S^{\rm dual}_{\rm eff}[\psi,A]=\int d^3 x \Bigg\{ \frac{1}{2}F^{\mu}F_{\mu} - \frac{2\pi s_m}{ e_\chi^2}\epsilon^{\mu\nu\rho}A_\mu \partial_\nu A_\rho\\
+\bar\psi\Big[\gamma^\mu \left( i\partial_\mu + e_\psi F_\mu\right)+m\Big]\psi 
 \Bigg\}.
\eeq
In other words, the Pauli term couples the magnetic moment of the fermions with the magnetic field \cite{Jain:1989tx,Carrington:1994km,Nobre:1999mj}. Note that in $2+1$ dimensions the magnetic moment is a scalar leading to the coupling term $e_\psi k^\mu F_\mu$ seen in (\ref{Seff21}).

To analyse the properties of \eqref{Seff21} we integrate out the fermions $\psi$. As shown in \cite{Stern:1991fm,Georgelin:1991bh,Itzhaki:2002rc} the Pauli term can be obtained starting from the standard minimal coupling in the Dirac action $\bar\psi \gamma^\mu A_\mu \psi$ and shifting the gauge field $A_\mu$ into the generalised connection $A_\mu \rightarrow  A_\mu + e_\psi F_\mu$. We can then integrate out $\psi$ in \eqref{Seff21} by using the result \eqref{logdet} for the generalised connection and then set $A_\mu=0$, i.e
\beq\label{logdet2}
\Gamma[A_\mu + e_\psi F_\mu]\Bigg|_{A_\mu=0}=-i\log\left[\det\left( \gamma^\mu (i\partial_\mu+ A_\mu + e_\psi F_\mu)+m\right)\right] \Bigg|_{A_\mu=0} \approx \frac{  s_m  e_\psi^2}{8\pi}\int d^3 x\, \epsilon^{\mu\nu\rho}F_\mu \partial _\nu F_\rho.
\eeq
By using this result, which is also compatible with Ref. \cite{Dalmazi:2004vc}, the corresponding effective action takes the form
\beq\label{Seff2intpsi}
\bal
S^{\rm dual}_{\rm eff}[A]&=\int d^3 x \Bigg[\frac{1}{2}F^{\mu}F_{\mu} -\frac{2\pi s_m}{e_\chi^2}\epsilon^{\mu\nu\rho}A_\mu \partial_\nu A_\rho
+\frac{ s_m e_\psi^2}{8\pi}\epsilon^{\mu\nu\rho}F_{\mu}\partial _\nu F_{\rho}\Bigg].
\eal
\eeq
It is important to remark that, even though the higher-derivative term in the above action looks like a Chern-Simons form, it is not topological as it depends on the space-time metric. Indeed, as shown in \cite{Deser:1999pa}, up to boundary terms one can write
\beq
\epsilon^{\mu\nu\rho} F_\mu\partial_\nu F_\rho=\epsilon^{\mu\nu\rho} \Box A_\mu\partial_\nu A_\rho.
\eeq
Thus, this term leads to a non-vanishing contribution to the energy-momentum tensor, which is a signature of its non-topological nature.

As it has been shown in \cite{Deser:1999pa}, the action \eqref{Seff2intpsi} includes a ghost mode. Now we will show that this model admits a description in which the ghost is decoupled from the physical degree of freedom. In order to do so, we follow \cite{Bergshoeff:2009tb} and decompose the vector potential in terms of new variables $X$ and $Y$ as follows
\beq
A_0=\frac{1}{\sqrt{-\nabla^2}} X, \hspace{0.3cm}A_i=\frac{1}{\sqrt{-\nabla^2}} \varepsilon_{ij} \partial_j Y. \hspace{0.3cm}
\eeq
The effective action (\ref{Seff2intpsi}) then becomes
\beq\label{SeffXY}
\bal
S^{\rm dual}_{\rm eff}[X,Y]&=\int d^{3}x\Bigg[\frac{1}{2}Y\Box Y+\frac{ s_m e_\psi^2}{8\pi} X\Box Y-\frac{2\pi s_m}{e_\chi^2} XY+\frac{1}{2}X^{2}\Bigg].
\eal
\eeq
We can now integrate out the field $X$ in the corresponding partition function $Z=\int \mathcal D X \mathcal D Y e^{S_{\rm eff}^{\rm dual}[X,Y]}$, which yields an effective action for $Y$ given by
\beq\label{solX}
S_{\rm eff}^{\rm dual}[Y]=-\frac{1}{2}\left(\frac{e_\psi^2}{8\pi}\right)^2\int d^3x Y\left(\Box-m_+^{2}\right)\left(\Box-m_-^{2}\right)Y
\,,\hskip.5truecm
m_\pm^2=\frac{1}{2}\left(\frac{ 8\pi}{e_\psi^2}\right)^2\left(1 + \frac{e_\psi^2}{2e_\chi^2} \pm \sqrt{1 + 
\frac{e_\psi^2}{e_\chi^2}}\right).
\eeq
This higher derivative scalar field action can be expressed in terms of two Klein-Gordon fields $\varphi_\pm$ defined by
\beq
\label{m1m2}
\varphi_\pm=\frac{e_\psi^2}{8\pi}\frac{\left(\Box-m_\mp^{2}\right)Y}{\sqrt{|m_+^2-m_-^2|}}.
\eeq
The action then takes the form \cite{Hawking:2001yt,Kaparulin:2014vpa}
\beq
S^{\rm dual}_{\rm eff}[\varphi_+,\varphi_-]=\int d^3 x\left[ \frac{1}{2} \varphi_+\left(\Box-m_+^2\right)\varphi_+-\frac{1}{2}\varphi_-\left(\Box-m_-^2\right)\varphi_-\right].
\eeq
Hence, the field redefinition \eqref{m1m2} allows us to express \eqref{Seff2intpsi} as the action for two decoupled massive Klein-Gordon fields, $\varphi_+$ and $\varphi_-$. The field $\varphi_+$ is a physical Klein-Gordon field, while $\varphi_-$ is a ghost. Since the ghost fields is totally decoupled from the physical degree of freedom, the physical spectrum is not affected by it. In this sense we have a ``good'' ghost \cite{Hawking:2001yt} emerging in our theory. From \eqref{m1m2} we see that $m_+^2>0$ for any values of $e_\psi$ and $e_\chi$. On the other hand, $m_-^2$ can be positive or negative depending on the values of the couplings $e_\psi$ and $e_\chi$, implying that the ghost $\varphi_2$ can be also a tachyon. Thus, this theory shares similar features with the Chern-Simons-Maxwell theory \cite{Deser} that describes a single propagating massive bosonic mode. In our case, the effect of the higher-derivative term is to renormalise the topological mass of the boson. 

\section{Single and mutual Chern-Simons theories}

In this section we show that, besides the Chern-Simons and Maxwell terms, suitable choices of the parameters in the starting action \eqref{Action} lead to an effective description of the system that includes a mutual Chern-Simons term \cite{Birmingham}.

\subsection{Single Chern-Simons theory: the $\alpha=0$ case }

The case $\alpha=0$ corresponds to interaction couplings in \eqref{Action} that satisfy $V_\chi V_\psi =V_{\chi\psi}^2$. In this case we can define the four-spinor $\Psi=(\psi, \chi)^T$ and the corresponding current $J^\mu=\bar\Psi\Gamma^\mu \Psi$, where $\Gamma^\mu= \mathbb{I}_{2\times2} \otimes \gamma^\mu$. The generating functional \eqref{Zeta} then boils down to
\beq\label{Zalpha0}
\bal
Z=\int \mathcal  \mathcal D\bar\Psi \mathcal D\Psi
\,\exp\bigg\{i\int d^3 x\bigg[ 
\bar\Psi\left(i \Gamma^\mu \partial_\mu+m\right)\Psi 
+\frac{ 1}{2} e_\Psi^2 J^\mu  J_\mu  
\bigg]\Bigg\},
\eal
\eeq
where $e_\Psi^2 =e_\chi^2+e_\psi^2$. Since this is a standard Thirring model for $\Psi$ we can linearise the interactions by introducing a vector field $a_\mu$ \cite{Fradkin:1994tt}, so that by means of Gaussian integration we implement the replacement $e_\Psi^2 J^\mu J_\mu\rightarrow - \frac{1}{2}a^\mu a_\mu +   e_\psi a^\mu J^\mu$ in \eqref{Zalpha0}.
Using \eqref{logdet} to integrate out $\Psi$, the low energy behaviour of this system is captured by the following effective action
\beq
\label{CS1}
S_{\rm eff}[\psi,a]=\int d^3 x \Bigg[-\frac{1}{2}a^\mu a_\mu
+ \frac{s_me_\Psi^2}{8\pi} \epsilon^{\mu\nu\rho}a_\mu \partial _\nu a_\rho  
 \Bigg].
\eeq
This result can be also obtained from \eqref{Seff} by setting $\alpha=0$ in  and subsequently integrating out $\psi$. Following \cite{Deser:1984kw,Townsend:1983xs}, we can dualise this action to a Chern-Simons-Maxwell theory
\beq\label{MaxwelCS}
\bal
S^{\rm dual}_{\rm eff}[A]&=\int d^3 x \left[-\frac{1}{4}F_{\mu\nu}F^{\mu\nu}-M_A\epsilon^{\mu\nu\rho}A_\mu \partial_\nu A_\rho\right].
\eal
\eeq
Hence, for the specific case where $\alpha=0$ the system becomes formally equivalent to a single species self-interacting fermion that gives rise to a Chern-Simons theory with coupling
$M_A=2\pi s_m/(e_\chi^2+e_\psi^2)$. This theory describes massive bosons that only mediate short-range interactions \cite{Deser}.

\subsection{Mutual Chern-Simons theories: the $\xi=1$, $\alpha \neq 0$ case }

The choice of parameters $\xi=1$ and $\alpha \neq 0$ corresponds to the action \eqref{Action} with $V_\psi V_\chi>V_{\chi\psi}^2$. In this case the effective action \eqref{Seff} becomes
\beq\label{Seff3}
S_{\rm eff}[\psi,a]=\int d^3 x \Bigg[-\frac{1}{2}a^\mu a_\mu
+\frac{ s_m e_\chi^2}{8\pi} \epsilon^{\mu\nu\rho}a_\mu \partial _\nu a_\rho  
+\bar\psi\left(\gamma^\mu (i\partial_\mu+ e_\psi a_\mu)+m\right)\psi
+\frac{ \alpha^2}{2}j^\mu j_\mu \Bigg],
\eeq
so one can integrate out $\psi$ directly. Following similar steps as above, we use \eqref{gaussint} to linearise the self interaction in the path integral associated to \eqref{Seff3} by introducing a new vector field $b_\mu$. Subsequently, using \eqref{logdet} with the replacement $a_\mu \rightarrow a_\mu+\frac{\alpha}{e_\psi } b_\mu$ leads to
\beq\label{Seff2}
\bal
 S_{\rm eff}[a,b]&=\int d^3 x \Bigg[- \frac{1}{2} a^\mu a_\mu -\frac{1}{2} b^\mu b_\mu
+\frac{ s_m (e_\chi^2+e_\psi^2)}{8\pi} \epsilon^{\mu\nu\rho}a_\mu \partial _\nu a_\rho  
  +\frac{s_m \alpha^2}{8\pi} \epsilon^{\mu\nu\rho}b_\mu \partial _\nu b_\rho  
  + \frac{s_m e_\psi \alpha}{4\pi} \epsilon^{\mu\nu\rho}a_\mu \partial _\nu b_\rho
\Bigg].
\eal
\eeq
In this action both $a_\mu$ and $b_\mu$ are vector fields. In order to turn them into gauge fields we employ the interpolating action procedure. Consider the interpolating path integral (see appendix A in supplementary information)
\beq\label{Zint}
\bal
Z_I&=\int \mathcal D A \mathcal D B \mathcal D a \mathcal D b \,\exp\Bigg\{i\int d^3 x \Bigg[-
\frac{1}{2}a^\mu a_\mu- \frac{1}{2}b^\mu b_\mu +\epsilon^{\mu\nu\rho}a_\mu\partial_\nu A_\rho 
+\epsilon^{\mu\nu\rho}b_\mu\partial_\nu B_\rho
\\
&\hskip 5.2truecm-\frac{ m_A}{2} \epsilon^{\mu\nu\rho}A_\mu\partial_\nu A_\rho + m_I \epsilon^{\mu\nu\rho}A_\mu\partial_\nu B_\rho-\frac{ m_B}{2} \epsilon^{\mu\nu\rho}B_\mu\partial_\nu B_\rho  \Bigg]
\Bigg\},
\eal
\eeq
where the masses $m_A$, $m_B$ and $m_I$ are to be fixed in term of the couplings constants in \eqref{Seff2}. Integrating out the fields $A_\mu$ and $B_\mu$ leads exactly to the functional integral of the action \eqref{Seff2}, i.e.
\beq
Z_I= \int \mathcal D a \mathcal D b \,\exp\{i S_{\rm eff}[a,b]
\}\,,
\eeq
provided the masses $m_A$, $m_B$ and $m_I$ are given by
\beq\label{ms}
m_A
=\frac{4\pi s_m}{e_\chi^2}\,,\hskip1.2truecm
m_B
=\frac{4\pi s_m}{ \alpha^2}
\left(1+\frac{e_\psi^2}{ e_\chi^2}\right)
\,,\hskip1.2truecm
m_I
= \frac{4\pi s_m }{\alpha e_\chi^2} e_\psi\,.
\eeq
The dual theory is obtained by integrating out the vector fields $a_\mu$ and $b_\mu$ in \eqref{Zint}, which yields
\beq\label{ZeffAB}
\bal
Z_I&= \int \mathcal D A \mathcal D B \,\exp\Bigg\{i\int d^3 x \Bigg[
  -\frac{1}{4}F_{\mu\nu}F^{\mu\nu} 
-\frac{1}{4}G_{\mu\nu}G^{\mu\nu} 
\\
&\hskip4truecm -\frac{ m_A}{2} \epsilon^{\mu\nu\rho}A_\mu\partial_\nu A_\rho + m_I \epsilon^{\mu\nu\rho}A_\mu\partial_\nu B_\rho
-\frac{ m_B}{2} \epsilon^{\mu\nu\rho}B_\mu\partial_\nu B_\rho  \Bigg]
\Bigg\},
\eal
\eeq
where we have introduced a second field strength, $G_{\mu\nu}=\partial_\mu B_\nu -\partial_\nu B_\mu$.
Interestingly, for $\alpha^2=e^2_\chi+e^2_\psi$, there appears an emergent $\mathbb{Z}_2$ symmetry that exchanges the gauge fields, i.e.
\beq\label{Z2}
\mathbb{Z}_2: \hspace{0.2cm} A_\mu\rightarrow B_\mu,\hspace{0.2cm} B_\mu\rightarrow A_\mu.
\eeq
 This theory describes two massive bosons and generalises the Chern-Simons-Maxwell theory \cite{Deser}, which is defined for a single $U(1)$ gauge field and the double-Maxwell-BF theory \cite{Mavromatos,Hansson,Palumbo}. The latter, defined for $m_A=m_B=0$, has been employed to study the Meissner effect in two-dimensional superconductors/superfluids that preserve time-reversal symmetry. In this context, the two massive bosons can be interpreted as massive modes related to an effective London penetration length \cite{Hansson}.\\
 Here, we give a physical interpretation of our model by neglecting the Maxwell terms and focusing on the topological sector 
\beq\label{CStermA}
S_{\rm eff}^{\rm dual}[A,B]=-\int d^3x \epsilon^{\mu\nu\rho}\Bigg[ \frac{ m_A}{2} A_\mu\partial_\nu A_\rho -  m_I A_\mu\partial_\nu B_\rho
+\frac{ m_B}{2} B_\mu\partial_\nu B_\rho\Bigg],
\eeq
which is dominant at large distances.
By a suitable rescaling of the gauge fields, this topological action formally coincides with that one derived in Ref.~\cite{Moore} in thin-film topological insulators. In this context, our $T$-broken action would describe an emergent quantum anomalous Hall state induced by interactions. In fact, the presence of $s_m$ in all the three coefficients $m_A$, $m_B$ and $m_I$ is the signature of the presence of a common Chern number encoded in those terms that changes sign when the external Zeeman field is flipped. There are however important physical differences with respect to Ref. \cite{Moore}. In that work, the $A_\mu$ field is identified with an external electromagnetic field and the two fermion species live on different boundaries, such that only in the thin-film limit the effective 2D model for the boundary contain both species.

Finally, note that the effective action in (\ref{CStermA}) can be further reduced by integrating out the gauge field $B_\mu$, which yields the Chern-Simons action we met in \eqref{MaxwelCS} with the same mass $M_A=2\pi s_m/(e_\chi^2+e_\psi^2)$. 
Therefore, at the level of the topological affective action, integrating out $B_\mu$ is equivalent to set $\alpha=0$ in the original action \eqref{Action}. 
On the other hand, if we choose to integrate the gauge field $A_\mu$, we obtain the Chern-Simons term of \eqref{MaxwelCS} for the field $B_\mu$ with $M_B=2\pi s_m/\alpha ^2$. 
This result corresponds to setting $e_\psi=e_\chi=0$ in \eqref{Action}, which eliminates the interaction between $\psi$ and $\chi$ and keeping only the Thirring self-interaction for $\psi$ with coupling $\alpha^2$.

\subsection{Response to external electromagnetic field}

In this subsection we probe the system in Eq.~\eqref{Seff3} by introducing an external electromagnetic potential $\mathbb A_\mu$. This requires to modify action \eqref{Action} by minimally coupling the fermions $\psi$ and $\chi$ to $\mathbb A_\mu$ as follows
\beq
S[\chi,\psi, \mathbb A]= S[\chi,\psi] +q \int d^3 x (j^\mu +k^\mu)\mathbb A_\mu.
\eeq
Repeating the steps outlined in the previous section leads to the generalization of the effective action \eqref{Seff2} to the following one
\beq
 S_{\rm eff}[a,b,\mathbb A]=  S_{\rm eff}[a,b] +\frac{s_m q}{4\pi}\int d^3x \epsilon^{\mu\nu\rho}
 \left[
 q \mathbb A_\mu \partial_\nu \mathbb A_\rho
  +(e_\chi+ e_\psi)\mathbb A_\mu \partial_\nu a_\rho
 +\alpha  \mathbb A_\mu\partial _\nu b_\rho\right].
\eeq
The interpolating path intgral \eqref{Zint} is then modified as
\beq\label{ZintExt}
\bal
Z_I&=\int \mathcal D A \mathcal D B \mathcal D a \mathcal D b \,\exp\Bigg\{i\int d^3 x \Bigg[-\frac{1}{2}a^\mu a_\mu- \frac{1}{2}b^\mu b_\mu +\epsilon^{\mu\nu\rho}a_\mu\partial_\nu A_\rho 
+\epsilon^{\mu\nu\rho}b_\mu\partial_\nu B_\rho
 \\
&
-\frac{ 2\pi s_m}{e_\chi^2} \epsilon^{\mu\nu\rho}
\left(
A_\mu\partial_\nu A_\rho 
-\frac{2 e_\psi}{\alpha } A_\mu\partial_\nu B_\rho
+\frac{e_\chi^2+ e_\psi^2 }{  \alpha^2} B_\mu\partial_\nu B_\rho 
\right) 
+\frac{q}{e_\chi} \epsilon^{\mu\nu\rho} \left(
\mathbb A_\mu \partial_\nu A_\rho
+\frac{e_\chi-e_\psi}{\alpha}  \mathbb A_\mu \partial_\nu B_\rho\right)
  \Bigg]
\Bigg\},
\eal
\eeq
and the dual action \eqref{CStermA} is generalized to
\beq\label{actionCSext}
S_{\rm eff}^{\rm dual}[A,B,\mathbb A] = S_{\rm eff}^{\rm dual}[A,B]
+
\frac{q}{e_\chi} \epsilon^{\mu\nu\rho} \mathbb A_\mu \partial_\nu A_\rho
+\frac{q}{\alpha} \left(1-\frac{e_\psi}{e_\chi}\right)\epsilon^{\mu\nu\rho} \mathbb A_\mu \partial_\nu B_\rho.
\eeq
Integrating out the fields $A_\mu$ and $B_\mu$ in the corresponding path integral reduces \eqref{actionCSext} to the following action
\beq\label{Lep2hoIII}
\tilde{ S}_{\rm eff}^{\rm dual}[\mathbb A] =\frac{ q^2 s_m}{4\pi}\int d^3 x  \epsilon^{\mu\nu\rho}\mathbb A_\mu\partial_\nu \mathbb A_\rho,
\eeq
which is given only in terms of $\mathbb A_\mu$.
Varying this action with respect to $\mathbb A_\mu$ we obtain the current
\beq\label{current1h0}
\mathbb J^\mu=\frac{\delta \tilde{ S}_{\rm eff}^{\rm dual}}{\delta \mathbb A_\mu} =\frac{q^2 s_m}{4\pi}\epsilon^{\mu\nu\rho} \mathbb F_{\nu\rho}.
\eeq
This clearly shows a topological Hall response of our system in presence of an external electromagnetic field. This current is sensitive to the sign $s_m$ of the generated mass $m$, but it is insensitive to the particular values of the interaction couplings, $e_\chi$, $e_\psi$ and $\alpha$.

\subsection{Domain walls and chiral bosons}

We now show that our effective topological field theory in Eq. \eqref{CStermA} allows us to describe the 1D gapless modes trapped along defect lines (namely, 1D domain walls) that we can add on the 2D gapped boundary.
In fact, defect lines behave as an effective spacial boundary for the 2+1-D bosonic model in Eq. 
\eqref{CStermA} and the CS/CFT correspondence \cite{Fradkin} allows us to derive the chiral boson action associated to the 1D modes \cite{Wen}.
For this action, we can define the following new fields
\beq
A^\pm= A\pm \frac{\sqrt{ e_\chi^2 + e_\psi^2}}{\alpha} B.
\eeq
In this way, the effective action takes the form of two decoupled Chern-Simons terms
\beq\label{chiralaction}
\tilde{ S}_{\rm eff}^{\rm dual}
= \int d^3 x \epsilon^{\mu\nu\rho} \Bigg[
\kappa_+ A^+_\mu \partial_\nu A^+_\rho 
+
\kappa_- A^-_\mu \partial_\nu A^-_\rho
\Bigg],
\eeq
where we have defined
\beq
\kappa_\pm =-\frac{\pi  s_m}{ e_\chi^2}
\left(1\pm
 \frac{e_\psi  }{\sqrt{ e_\chi^2  + e_\psi^2 }  }
 \right).
\eeq
Following \cite{Fradkin}, we adopt coordinates $(t, x,  y)$ and consider the generalized axial gauge
\beq
A^\pm_{ t}- v A^\pm_{x}=0.
\eeq
The Gauss law $F_{xy}=0$ leads to locally pure gauge configurations
\beq
A^\pm_{x} =\partial_x \Phi^\pm \,,\qquad A^\pm_{y} =\partial_y \Phi^\pm,
\eeq
which can be implemented directly in the action \eqref{chiralaction} and leads to
\beq\label{edg2.2}
\bal
S_{\rm edge}
=  \int dt dx \bigg[\kappa_+ \left(\partial_{ t} \Phi^+ \partial_{ x} \Phi^+ -v (\partial_{ x} \Phi^+ )^2 \right)+ \kappa_- \left( \partial_{ t} \Phi^ - \partial_{ x} \Phi^-
-v  (\partial_{ x} \Phi^- )^2\right) \bigg].
\eal
\eeq
Therefore, the 1D dynamics is described by two chiral bosons, which are determined by the parameters $\kappa_\pm$ and velocity $v$ \cite{Wen}.
Importantly, these chiral modes trapped along the line defects can be eventually measured in experiments.

\section{Conclusions}

In this article we have studied the effect interactions have on two Dirac fermions in $2+1$ dimensions. As we are interested in the topological properties of this system we employed the bosonisation method in order to obtain the corresponding effective gauge theories. As we vary the fermion couplings with intra-species interactions, $V_\chi$ and $V_\psi$, and inter-species interactions $V_{\chi\psi}$ we obtain a variety of topological theories that correspond to different phases of the model. When one of the fermionic species does not self-interact, $V_\psi=0$, then the system is described by a Chern-Simons theory with a higher-derivative term. With the appropriate field reparametrisation this theory can be written in terms of a physical scalar field and a ``good ghost" that completely decouples from the physical spectrum. Hence, it gives a well behaved topological theory, which shares similar features with the Chern-Simons-Maxwell theory and the topological mass is renormalised by the higher-derivative term.
Beyond this particular regime, when we take $V_\chi V_\psi>V_{\chi\psi}^2$ the action is given in terms of two coupled Chern-Simons theories that describes two propagating massive bosons. In this case the system is described by an emergent quantum anomalous Hall state induced by interactions and the two interacting massive Dirac fermions can be mapped to the two massive bosons. Moreover, for a particular choice of the coupling constants, there appears an emergent $\mathbb{Z}_2$ symmetry.
In terms of physical observables, we have shown that by coupling the interacting model to an external electromagnetic field, the semiclassical currents are related to a topological Hall response. Moreover, by adding suitable domain walls on the gapped boundary, there appear propagating 1D modes trapped along the domain walls (i.e. defect lines). This is due to the well-known CS/CFT correspondence, where the CFT describes the 1D chiral modes, which can be in principle measured in experiments.
Our method does not have a simple interpretation in the case where $V_\chi V_\psi<V_{\chi\psi}^2$ so an alternative approach needs to be taken. We leave this case for a future investigation. Finally, note that our approach can be naturally generalised in various ways. One can consider multi-species interactions described by multi-U(1) gauge fields. This paves the way to study the interacting boundaries of 3D topological crystalline insulators for $n_M>2$ through functional bosonisation. Moreover, one can consider multi-SU$(N)$ non-Abelian generalisation of the gauge fields along the lines of Ref.~\cite{Palumbo:2013oba}.

\acknowledgments

We would like to thank J. Gomis for useful discussions. This work was supported by the ERC through the Starting Grant project TopoCold and the EPSRC grant EP/R020612/1. Statement of compliance with EPSRC policy framework on research data: This publication is theoretical work that does not require supporting research data. PS-R acknowledges the School of Physics and Astronomy of the University of Leeds for hospitality and support as invited researcher.

\providecommand{\href}[2]{#2}\begingroup\raggedright\endgroup

\end{document}